\documentclass[11pt]{article}
\pdfoutput=1
\usepackage{putex}
\usepackage{amsmath,hyperref,amssymb,amsfonts,bm,amscd,slashed,ytableau}
\usepackage{cite}
\usepackage{graphicx}
\usepackage{multirow}
\usepackage{verbatim}
\usepackage{appendix}
\usepackage{fancybox}
\usepackage{array}
\usepackage{url}
\usepackage{float}
\usepackage{mathtools}
\usepackage{mathrsfs}
\usepackage{bbm}
\usepackage[bbgreekl]{mathbbol}

\DeclareSymbolFontAlphabet{\mathbbm}{bbold}
\DeclareSymbolFontAlphabet{\mathbb}{AMSb}

\DeclareMathAlphabet{\mathpzc}{OT1}{pzc}{m}{it}

\newcommand{\bea}{\begin{eqnarray}}
\newcommand{\eea}{\end{eqnarray}}
\newcommand{\be}{\begin{equation}}
\newcommand{\ee}{\end{equation}}

\def \beaa {\begin{equation}\begin{aligned}}
\def \eeaa {\end{aligned}\end{equation}}

\newcommand{\C}{{\mathbb C}}

\newcommand{\dd}{{\rm d}}

\def\cA{{\mathcal A}}

\def\cH{{\mathcal H}}

\def\cR{{\mathcal R}}
\def\cS{{\mathcal S}}

\def\cW{{\mathcal W}}

\numberwithin{equation}{section}
\begin{document}
%\preprint{}

\institution{SCGP}{Simons Center for Geometry and Physics,\cr Stony Brook University, Stony Brook, NY 11794-3636, USA}
	
\title{Snowmass White Paper:\\ The Quest to Define QFT}
\authors{Mykola Dedushenko\worksat{\SCGP}}
	
\abstract{This article provides a review of the literature on rigorous definitions and constructions in Quantum Field Theory, spanning the period of seven decades. Comparing with the ideas and constructions found in the modern physics literature, we conclude that none of the existing systems of QFT axioms can cover all the physical situations. Therefore, it is still an outstanding open problem to formulate a complete definition of QFT. We argue that the question is of relevance for both physicists and mathematicians.}
	
\date{}
	
\maketitle
	
%\tableofcontents
	
\section{Introduction}
The subject of Quantum Field Theory is nearing the centennial, with its inception dating back to the papers \cite{Born:1926uzf,Dirac:1927dy}, followed by \cite{Heisenberg:1929xj,Heisenberg:1930xk} and many others. Growing mostly out of the need to reconcile special relativity with quantum mechanics, both young subjects at that time, it led to the development of the early version of perturbative QFT over the following two decades. An interesting historical account of those early years can be found in the first chapter of Weinberg's excellent textbook \cite{Weinberg:1995mt}. One of the biggest challenges that had to be overcome were the UV divergences, which eventually led to the development of renormalization techniques by the end of forties in the works of Dyson, Feynman, Schwinger, and Tomonaga, \cite{Schwinger:1948iu,Schwinger:1948yj,Schwinger:1948yk,Schwinger:1949ra,Feynman:1948ur,Feynman:1948fi,Feynman:1948km,Tomonaga:1946zz,Koba:1947rzy,10.1143/ptp/2.4.198,10.1143/ptp/3.1.1,10.1143/ptp/3.3.290,Tomonaga:1948zz,Dyson:1949bp}.% with early contributions by Kramers and Bethe. 

A new field, with all its strange renormalization machinery, desperately needed a clear set of rules, or axioms, from which everything else would follow in a logical manner.  Such rules would ``distill'' the subject into a mathematical subfield, but they were also necessary due to the limitations of the perturbative Lagrangian techniques. Thus starting from the fifties, various axiomatics for QFT began to appear. For the purpose of this paper, we will consider Wightman's axioms \cite{Wightman:1956zz,zbMATH03147407,WightGard,Wightman:1963deu} as the start of that process, which will inevitably miss some earlier attempts, such as the S-matrix program of Heisenberg \cite{Heis1943}, a closely related extended (off-shell) S-matrix approach of Bogolyubov-Medvedev-Polivanov \cite{Bogolyubov:1990kw}, or axioms of Lehmann-Symanzik-Zimmermann \cite{Lehmann:1954rq,Lehmann:1957zz} (the LSZ reduction formula, however, became part of the standard QFT formalism). Some of the other approaches are still being developed nowadays. Here we would like to give a brief overview of these issues, and discuss the range of applicability of various axiom systems. A point that we want to make is that \emph{none} of the existing definitions covers the full range of notions of Quantum Field Theory that appears in the physics literature.

Having a rigorous system of axioms for a physics subfield has two philosophical motivations. On the one hand, it provides a starting point for mathematical investigations, sort of extracting the abstract truth from the messy reality.  On the other, it indicates that the physical understanding of the subject has matured enough. Indeed, most other physics subfield (all the ``non-quantum'' physics and the nonrelativistic quantum mechanics) have already undergone this process. The fact that QFT does not have (as we will see) one clear and universal set of axioms likely shows that the physical understanding is still lacking. Hence, we argue, it is a challenge both for physicists and for mathematicians to define QFT. Below we will provide a brief overview of the existing approaches.

\emph{Announcement:} comments are highly appreciated!
	
\section{Existing axiomatics} 

\subsection{Correlator-focused approaches}
\paragraph{Wightman axioms.}
One of the older axiom systems that remains relevant to date is that due to Wightman \cite{Wightman:1956zz,zbMATH03147407,WightGard,Wightman:1963deu} (see also books \cite{StreaterWightman+2016,bogoliubov1975introduction,Bogolyubov:1990kw}). The axioms view fields as operator-valued tempered distributions and formalize the notion of expectation values of their products (``Wightman functions''). One starts with the assumption ({W0}) of relativistic invariance (as in Wigner's classification \cite{Wigner:1939cj}\footnote{Wigner's classification, generally, serves as an excellent starting point for the \emph{perturbative} description of Poincare-invariant QFT in flat space, as in, e.g., \cite{Weinberg:1995mt}.}, see also \cite{Barg47,Barg54,book:GelfMinlSha,book:Ohnuki}): physical Hilbert space is a unitary representation of the Poincare group. This assumption is supplemented by the spectral condition (the energy-momentum spectrum lies in the closed upper light-cone) and the uniqueness and Poincare-invariance of the vacuum state $\Psi_0\in\cH$. The axiom 
{W1} states that there is a set of \emph{fields} $\varphi_i[f]$ given by tempered distributions\footnote{These distributions are defined on smooth Schwarz functions $f\in\cS$, which allows to avoid the UV issues.} valued in the operators defined on (and preserving) a dense subset $D$ (which includes the vacuum) of the Hilbert space $\cH$. The subset $D$ is assumed to be Poincare-invariant. Then {W2} states covariance of fields with respect to the Poincare group, and {W3} requires locality (also called microcausality) in the form of (anti)commutativity of spacelike-separated fields. A quantum field theory is said to satisfy {W0}--{W3}, and in addition obey cyclicity of the vacuum: The span of vactors of the form $\varphi_1[f_1]\dots \varphi_n[f_n]\Psi_0$ (for all possible $n$ and $f_i$) is dense in $\cH$. The latter condition guarantees that there are enough fields in the theory.

A number of important results follow from these axioms, e.g., various properties of the Wightman functions $\cW(x_1,\dots,x_n)=(\Psi_0, \varphi_1(x_1)\dots \varphi_n(x_n)\Psi_0)$, such as relativistic transformation, spectral condition, Hermiticity condition, local commutativity, positive definiteness, and cluster decomposition, as well as analytic continuation \cite{Kallen:1960zqa,WightCmplx}. Most of these were proven in \cite{Wightman:1956zz} and reviewed in \cite{StreaterWightman+2016}, along with the reconstruction theorem recovering the data of ({W0})--({W3}) from the Wightman functions (for the case of scalar fields only). Other important references (especially on cluster decomposition stating that $\cW(x_1,\dots x_n)$ tends to $\cW(x_1,\dots,x_k)\cW(x_{k+1},\dots,x_n)$ when the groups of points $\{x_1,\dots, x_k\}$ and $\{x_{k+1},\dots, x_n\}$ are infinitely separated) include \cite{ARAKI1960260,HepJoRueSte,HaaSchro62,Ruelle62,JostHepp,Araki:1962zhd}, see also \cite{Schmidt1956QuantentheorieDF} on reconstruction. Such classical results as CPT theorem \cite{Luders:1954zz,Pauli55} and spin-statistics connection \cite{Fierz:1939zz,Pauli:1940zz} also follow from the axioms \cite{Jost:1957zz,Dyson:1958ogi,Luders:1958zz,Burgoyne1958,DELLANTONIO1961153}. The significance of local algebras of observables $\mathfrak{A}(\Omega)$ associated to spacetime regions (which will feature prominently in the Haag-Kastler axioms below) was already discussed by Haag in connection to Wightman axioms in \cite{Haag59CNRS}. See \cite{JOST1961127,Ruelle62,Borchers:1962cbs} for their further properties, questions of irreducibility, and \cite{Reeh:1961ujh} for the famous Reeh-Schlieder theorem asserting the cyclicity of the vacuum with respect to $\mathfrak{A}(\Omega)$. For some other important aspects, such as Haag's theorem, Borchers classes of local fields, Haag-Ruelle theory etc, see \cite{Haag:1955ev,HallWight,Greenberg:1959zz,Federbush:1960zz,Borch60,Kamefuchi:1961sb,Epstein1963OnTB,JostBook}.

\paragraph{Axioms of Euclidean QFT.} The Osterwalder-Schrader (OS) axioms \cite{Osterwalder:1973dx,Osterwalder:1974tc,Glaser:1974hy}, as well as their modifications by Glimm-Jaffe (GJ) \cite{Glimm:1987ng} and the axioms of Nelson \cite{NELSON197397} provide, roughly, the Euclidean version\footnote{See  \cite{Schwinger:1958qau,PhysRev.115.721,10.1143/PTP.21.241,Symanzik:1964zz,doi:10.1063/1.1704960,SymVare68} as well as references in \cite{JAFFE198531} for origins of the Euclidean QFT.} of Wightman axioms. OS axioms are also based on formalizing the notion of correlation functions, known as Schwinger functions $S_n(x_1,\dots, x_n) = \langle \phi_1(x_1)\dots \phi_n(x_n)\rangle$ in the Euclidean case.  They include: {OS0} temperedness of $S_n$ as distributions; {OS1} Euclidean covariance; {OS2} Reflection positivity; {OS3} (anti)symmetry under permutations (``anti'' in the fermionic case); {OS4} cluster decomposition. Under a subtle additional property of linear growth condition, the OS theorem \cite{Osterwalder:1974tc} (sometimes called the OS reconstruction theorem) states that the Schwinger functions can be analytically continued to the Minkowski signature to obey Wightman axioms there (see also modification by Zinoviev \cite{Zinoviev95}). The Glimm-Jaffe axioms similarly formalize the generating functional $S[f] = \left\langle e^{\phi[f]} \right\rangle \equiv \int e^{\phi[f]}\dd \mu$, where $\dd\mu$ is a measure on the space of distributions $\phi$. They demand its analyticity, regularity (in the form of a certain growth bound), Euclidean covariance, reflection positivity, and ergodicity of the time translations. These axioms imply OS with the growth condition, and thus also Wightman axioms. Finally, Nelson axioms \cite{NELSON197397,Nelson73,NELSON197397,NELSON1973211,Nelson:1973rp} similarly take the measure-theoretic approach seriously and, importantly, require \emph{Markov property}, which essentially captures locality and implies that the state of a system in some region is assigned to the boundary of this region (and in particular, the Hilbert space is assigned to the boundary). He also requires ergodicity and proves that the Wightman axioms follow upon analytic continuation to the Minkowsi space (``Nelson's reconstruction theorem'', see book \cite{Simon:1974dg} by B.Simon for the review of this and other topics; see also \cite{Hegerfeldt1974FromET}). Also note the result of \cite{FROHLICH1980237} extending the OS reconstruction to equilibrium statistical mechanics (see also \cite{BirFro}), and the result of \cite{FroOstSei} studying the reconstruction of representations of the Poincare group in the same context of Euclidean QFT.
A recent work \cite{Lee:2021zsc} also provides generalizations of W and OS axioms (and reconstruction theorems) that are supposed to be suitable for gauge theories.

\paragraph{Constructive Field Theory.} The axioms discussed so far are completely non-constructive, one has to do extra work to provide examples. At first, only free fields (including generalized free fields defined in \cite{Greenberg:1961mr},) and various solvable models related to free fields were known to satisfy the Wightman axioms (and, naturally, other axiom systems). This led to the subject of \emph{Constructive Field Theory} (CQFT) emerging in the 1960's \cite{Jaffe66,lanford_1966,Nelson_conf66,Glimm:1968kh,Jaffe:1969wf}, whose main goal was to provide rigorous interacting examples of QFTs. An extensive review (as of 1987) can be found in the book \cite{Glimm:1987ng} (it focuses on the Euclidean path integral approach, see also other books on the subject: \cite{FuncIntSimon,bookPI2011,MathTheorPathInt,MathPathInt}), also see \cite{Jaffe:2006uz,Jaffe:2007uy,rivasseau2014perturbative, BaezSegalZhou, ConstrPhys}. An early vision of the field can be found in \cite{Wightman:1972ru,cqftVeloWight}, as well as a slightly later review \cite{Balaban:1986cv}; a review summarizing some successes of CQFT as of 2000 is in \cite{Jaffe:2000ub}, as well as a slightly more detailed review in \cite{Rivass2000}. More recent reviews include \cite{Jaffe07Rev,summers2016perspective,SummWeb} and a talk \cite{JaffeTalk2020}. Through a lot of work starting from the late sixties, success has been achieved in rigorously constructing and studying 2d scalar theories with arbitrary polynomial interaction (the so-called $\mathcal{P}(\phi)_2$ theories) \cite{Glimm:1968kh,GlimmJa:1970,GlJa70.3,Glimm:1972kn,Cannon:1970wp,PhysRevLett.28.1213,10.2307/1970988,Glimm:1973ga,Glimm:1974tt,10.2307/1970959,Spencer:1974uw,Spencer:1975hk,Fro74P2,Frohlich:1974ejk,Dimock:1976er,GlimmJafSpe76,FroSim77,Glimm:1975tw,Summers:1979pc,Haba:1979bj,DYNKIN1983167} (see also book \cite{Simon:1974dg}, recent papers \cite{Shen:2021nww,Frohlich:2022itn}, and \cite{Frohlich:1977yy,Park:1977fg,Osipov:1981un,Barashkov:2021oup} on other potentials), three-dimensional $\lambda \phi^4$ theory \cite{Glimm:1973kp,Feldman:1976im,Magnen:1976fw,Glimm:1976tr,Glimm:1976ts,Frohlich:1976xk,Magnen:1977ha,Park:1977gq,BryFroSok83} (see also more recent \cite{Gubinelli:2021nou,Albeverio:2021ewk,Hairer:2021byp,Oh:2021zvt,Jagannath:2021ewg}), Gross-Neveu theory \cite{Gawedzki:1985uq,Gawedzki:1985ez,Feldman:1985ar,Feldman:1986ax,Disertori:1998qe,SalWiecz}, Thirring model \cite{Frohlich:1976mt,Benfatto:2006ey,Benfatto:2007qp,Mastropietro:1993mf,Falco:2005pgf} (in particular, all these theories were shown to obey Wightman axioms);
other theories with fermions include 2D and 3D Yukawa models \cite{Osterwalder:1972vwp,Schrader:1972yr,Osterwalder:1973zr,Osterwalder:1973dx,Glimm:1970ze,Glimm:1971sjc,Seiler:1974ne,Seiler:1975aj,Seiler:1975zt,Seiler:1976yk,McBryan:1973sc,Mcbryan:1975wh,McBryan:1975gx,McBryan:1975yk,McBryan:1975pp,McBrPark,Magnen:1976uc,Cooper:1977jum,Osipov:1977pc,Osipov:1980nn,Renouard:1977wr,Renouard:1979yu,MagSen80,Balaban:1980er,Lesniewski:1987ha,Frohlich:1974zs} (see also \cite{Frohlich:1987gu}) and some supersymmetric models \cite{Nicolai:1978ic,Nicolai:1978ug,Jaffe:1987du,Jaffe:1987ys,Jaffe:1987fj,Jaffe:1988fh,Janowsky:1990jm}. Random walks representations of Euclidean theories were introduced \cite{Symanzik:1964zz,SymVare68} and developed later \cite{BryFroSpe82,BryFrohSok83,Frohlich:1984gev,Aizenman1985}, resulting in various applications \cite{BryFroSok83,Chayes:1985fw}, most prominently the proof of triviality of the $\phi^4$ theory in $d\geq 5$ spacetime dimensions \cite{Aizenman:1981du,Aizenman:1982ze,Frohlich:1982tw} (see book \cite{RandWalkBook}). The four-dimensional case turned out to be much more subtle \cite{Sokal:1981dx,Gawedzki:1982ub,AragaoDeCarvalho:1983mua,AIZENMAN1983261,Gawedzki:1985ic,Gawedzki:1985cf,GawComm85,Hara1987ARC,FeMagRivSen87,Bauerschmidt:2014tba} and has been resolved only recently in \cite{Aizenman:2019yuo} (see also lecture notes \cite{Aizenman:2021yve}) confirming triviality of the $\phi^4_4$ model.\footnote{``Trivial'' means ``free'' or ``Gaussian'', and the statements are about UV-complete models (i.e., with cutoffs removed) in precisely integral dimensions. Of course nothing prevents models with cut-offs from being nontrivial effective field theories, and furthermore, $d=4-\epsilon$ \cite{Wilson:1971dc} is not covered by such statements.} Lattice regularization has played role, especially in gauge theories \cite{Wilson:1975id,Bell:1975wtp,Wilson:1977nj}, see for example \cite{Brothier:2019asa} and many references therein, especially works \cite{Balaban:1982nn,Balaban:1982ce,Balaban:1982xc,Balaban:1983dc,Balaban:1983bj,Balaban:1984qr,Balaban:1984qpa,Balaban:1984dg,Balaban:1984qn,Balaban:1983dd,Balaban:1984qp,Balaban:1985pi,Balaban:1985za,Balaban:1985yy,Balaban:1987yj,Balab88,Balaban:1988rr,Balaban:1989ax,Balaban:1989qz,Balaban:1989ay} of Balaban, see also \cite{Dimock:2011qf,Dimock:2012kx,Dimock:2013oga} that revisits Balaban's approach to the renormalization group (illustrated with the $\phi^4$ theory) and \cite{Dimock:2015tya,Dimock:2015cfs,Dimock:2017dmd}. The results of \cite{Balaban:1985pi,Balaban:1985yy} and \cite{Magnen:1992wv} (using different methods) provide a significant progress towards solving the Millennium problem on the four-dimensional Yang-Mills,\footnote{Curiously, after two decades of rapid progress in the 70s and 80s, the field of Constructive Field Theory has gone so far away from the mainstream that, even though it hosts one of the most famous problems in theoretical physics, many young people nowadays do not even know that this field exists. In authors opinion, this state of affairs will change in the future, as more mathematicians are starting to think about QFT nowadays again.} see \cite{Jaffe:2000ne,Doug2004} for discussions. 

\subsection{Algebraic QFT}
\paragraph{Haag-Kastler axioms.} 
Algebraic QFT (AQFT) is another approach to axiomatizing QFT that de-emphasizes the notion of fields, and instead formalizes the algebra of observables without referring to any Hilbert space at first. This subject was initiated by the formulation of Haag-Kastler (HK) axioms in \cite{Haag:1963dh} (with some elements appearing in earlier works, such as \cite{Haag59CNRS,HaaSchro62,Araki62Lect,Haag:1958vt,Ruelle62,Segal47,Segal63}, see also reference [2] in \cite{Haag:1963dh}). There exists a number of books and monographs on AQFT \cite{HaagBook,ArakiBook,Baumgaertel:1992vy,Baumgaertel:1995bb,Horuzhy,Emch72,Borchers:1996kn} and on operator algebras \cite{BraRob1,BraRob2,Dixm1,Dixm2,KadRing,Naimark72,Pedersen2018,Sakai1,Sakai2,SunderBook,StratZsi,Stratila2021,Takesaki1,Takesaki2,Takesaki3,Jones_vN,EvTak1,EvTak2,Inoue98}, which should be consulted for details. Among the more recent literature, we mention a collection \cite{Rome-2015}, a concise review \cite{Fewster:2019ixc}, books and monographs \cite{Halvorson:2006wj,Rejz-2016,Michae-2019} and a related book \cite{Ent-2018}. The key points and references are also summarized in \cite{HKnLab}. The HK axioms (sometimes called Araki-Haag-Kastler axioms) are about relativistic local unitary QFTs in flat Minkowski space-time. For every causally closed\footnote{See discussion of causal closedness in \cite{CausComplNLab}} subset $U$ of observables one assigns a $C^*$-algebra\footnote{$C^*$-algebras were introduced in the works of Gelfand and Naimark. See above block of references and \cite{ArakiBook} on $C^*$ algebras. In short, a $C^*$-algebra is characterized by the following data: a $\C$-algebra with an involution $*$ obeying natural properties (this is a \emph{$*$-algebra}); norm $\| A\|$ obeying $\|AB\|\leq \|A\| \|B\|$ and $\| A^*\| = \|A\|$; completeness with respect to the topology induced by $\|\cdot \|$ (Banach $*$-algebra, or $B^*$-algebra); the $C^*$-property $\| A^* A\| = \|A\| \|A^*\|$.} of observables $\mathfrak{A}(U)$. Under an inclusion $U_1\subset U_2$, one has an inclusion $\mathfrak{A}(U_1) \subset \mathfrak{A}(U_2)$ of $C^*$-algebras (this property is called isotony), which is functorial, i.e., respects compositions (this data is often called a \emph{local net of algebras}; we could say that $\mathfrak{A}(\cdot)$ is an isotonic pre-cosheaf, except it is defined on causally closed subsets rather than opens). The requirement of causal locality says that $\mathfrak{A}(U_1)$ and $\mathfrak{A}(U_2)$ commute with each other (inside $\mathfrak{A}(U)$, where $U_i\subset U$) if $U_1$ and $U_2$ are spacelike separated. Furthermore, one usually imposes Poincare covariance in the form of a morphism $\alpha(p): \mathfrak{A}(U) \to \mathfrak{A}(pU)$ for any Poincare transformation $p$. Another requirement, due to the existence of linear dynamics, known as a time slice axiom, states that if $U_1 \subset U_2$ and $U_1$ contains a Cauchy surface of $U_2$, then $\mathfrak{A}(U_1) \to \mathfrak{A}(U_2)$ is an isomorphism. One also often requires positivity of the energy spectrum (i.e., of the operator of time translations).

In the algebraic formulation, quantum states are understood as linear maps $\omega: \mathfrak{A} \to \C$ satisfying the positivity condition $\omega(A^* A)\geq 0$ for all $A\in\mathfrak{A}$, where $\omega(A)$ is the ``expectation value of $A$''. One can consider faithful representations $\pi_\omega$ of algebras $\mathfrak{A}$ by bounded operators $\pi_\omega(\mathfrak{A}(U))\subset{\mathfrak B}(\cH)$ on the Hilbert space $\cH$ obtained via the GNS construction from the state $\omega$ \cite{GelNeu43,SegalGNS}. The algebra ${\mathfrak B}(\cH)$ has two useful notions of closed $*$-subalgebras: the above-mentioned $C^*$-algebras (closed in the norm topology), and von Neumann algebras \cite{Neum30} (which are closed in either one of the three topologies: strong operator, weak $*$, and weak operator), see for example \cite{HaagBook} or any other textbook referred above. Sometimes $\mathfrak{A}(U)$ is taken to be a $C^*$-algebra, but quite often one focuses on $\cR(U)=(\mathfrak{A}(U))''$, which is the minimal von Neumann algebra containing $\mathfrak{A}(U)$, where $(\cdot)'$ denotes the commutant\footnote{In the subject of operator algebras, the ``commutant'' of $X$ means everything that commutes with $X$, while in the rest of math this notion would be called a centralizer.} inside ${\mathfrak B}(\cH)$. One often talks about the net $\cR(U)$ forming a \emph{vacuum representation}, see, e.g., a review \cite{NCatVacRep}. This then connects to the rich theory of von Neumann algebras \cite{MurrNeum1,MurrNeum2,vonNeumann1939,Neum3,MurrNeum4,Neum43,Neum49} (see collection \cite{NeumCollect} and textbooks cited previously), in particular such topics as: Decomposition \cite{Neum49} into factors of Type I, II, III \cite{MurrNeum1} depending on whether the spectrum of dimensions of projectors on the invariant subspaces in $\cH$ is, respectively, discrete containing all integers in an interval, continuous, or consisting of just $0$ and $\infty$,  and a deep result that in QFT we deal with the Type III factors \cite{Araki64_type,Longo82,Fredenhagen85}; modular or Tomita-Takesaki theory (introduced by Tomita \cite{Tomi67_prep} and clarified by Takesaki \cite{Takesaki:1970aki}, see also \cite{Takesaki2} and expositions by Borchers \cite{Borchers:2000pv} and Summers \cite{Summers:2003tf}, or the book \cite{HaagBook},) which provides the structural theory of the Type III factors, and has connections to other topics, such as KMS states (see \cite{Haag:1967sg,Takesaki:1973ge,Borchers:1998ye}). There has been a lot of interest in the modular theory recently due to its connection to the entanglement properties in QFT (see \cite{Witten:2018zxz} as the entrance point into this portion of literature).

Similar to other approaches, it is possible to study general structural properties within the AQFT system of axioms, such as existence of scattering states \cite{Haag:1958vt,Ruelle62}, superselection sectors\footnote{The concept of superselection sectors \cite{Wick:1952nb,Wick:1970bd}, as it is apparent from \cite{Haag:1963dh}, was from the beginning important in the development of AQFT, see also review \cite{Streater:1975vw}. The idea was that different superselection sectors arise from the inequivalent representations of one algebraic structure.} \cite{Doplicher:1971wk,Doplicher:1973at} (see also, e.g., \cite{Landsman:1989ka,Landsman:1990jt,Landsman:1990dw,Casini:2019kex}), spin-statistics and CPT theorems (see \cite{Guido:1995fy,Verch:2001bv,Borchers:2000fw}, also see the topological version in \cite{Johnson-Freyd:2015fua}), the Reeh-Schlieder theorem \cite{Reeh:1961ujh} (it was already mentioned earlier, but traditionally this theorem is viewed as part of the AQFT machinery), Goldstone theorem \cite{Kastler:1966wdu,Ezawa:1966zz}.

One difference from the Wightman axioms should be clear: while they did not require boundedness of the operators (e.g., the momentum operator has unbounded spectrum), this was sort of an idealization. Any realistic experiment involves devices with finite ranges of possible values, thus any outcome should be predictable with arbitrary precision by a theory dealing with bounded operators only, like in the AQFT framework.\footnote{The idea to replace unbounded operators by bounded, bypassing technical issues in quantum mechanics, goes back to works of von Neumann, and in the case of QFT I.Segal suggested to use the same method \cite{Segal47}.} In the discussion of connection to the Wightman axioms, one asks two questions: whether, starting with a Wightman field smeared out with a compactly supported test function, one can find a self-adjoint bounded operator, and whether, starting with a net of algebras of bounded operators, one can obtain Wightman fields by a limiting process shrinking the regions to points (such questions were studied, e.g., in \cite{BorZimm64,Driessler:1986pk,Borchers:1989az,Borchers:1991vy,Borchers:1992rq,Fredenhagen:1981zw,Summers:1987db}).

\paragraph{Perturbative AQFT.} Requirements of bounded, $C^*$ or von Neumann, are dropped in \emph{perturbative AQFT}, where instead one deals with formal series star-algebras. Reviews include \cite{Fredenhagen:2012sb,Fredenhagen:2015iia,Rejz-2016}, see also a book \cite{Dutsch:2019wxk} and expositions \cite{SchRev,NLabPertQFT}. A few references on causal perturbation theory relevant in this context are: \cite{Stueck51,StuAndRiv50,StueckelbergdeBreidenbach:1952pwl,Salam51,BogoPara57,EpsGlas73,Ilyin:1978gq}, books \cite{Scharf1,Scharf2,Scharf3} and a review \cite{Aste:2009mj}. A more recent block of papers on the formalism of perturbative AQFT is \cite{Duetsch:1998hf,Duetsch:2000nh,Duetsch:2002yp,Hollands2003ALGEBRAICAT,Brunetti:2004ic,Duetsch:2004dd,Dutsch:2005bm,Brennecke:2007uj,Keller:2010xq,Rejzner:2011au,Duetsch:2013xca,Crawford:2021adf,Crawford:2022ost} (including \cite{Hollands2003ALGEBRAICAT} on the $1/N$ expansion) and \cite{Brunetti:2009qc} (see also comments in \cite{Stora:2008rfm}). See also \cite{Fredenhagen:2011an,Fredenhagen2011BatalinVilkoviskyFI,Rejzner:2011jsa,Rejzner:2013ak,Rejzner:2020bsc} and \cite[Chapter 7]{Rejz-2016} on the role of Batalin-Vilkovisky formalism \cite{Batalin:1981jr,Batalin:1983ggl}, and \cite{Dito:1990rj,Duetsch:2000de,Hirshfeld:2002uu,2016arXiv160309626C,Hawkins2016TheSP} on the relation to deformation quantization.

\paragraph{AQFT in curved space.} Quantum field theory on curved space pushes the limits of applicability of the QFT machinery.\footnote{Here we set the Newton constant to $0$, so the background is classical and non-dynamical.} It comes with new physical phenomena (such as particle production effects \cite{Parker:1969au}; Hawking effect \cite{Hawking:1975vcx}, with earlier precursors \cite{Zeldovich:1971mw,Starobinsky:1973aij,Unruh:1974bw}; Fulling-Davies-Unruh effect \cite{Fulling:1972md,Davies:1974th,Unruh:1976db}). They generally follow from the absence of Poincare invariance, and, as a result, absence of the distinguished vacuum, no particle interpretation, no momentum space representation, etc. Continuation from Mikowskian to Euclidean signature is also not generically available, and, relatedly, there is no unique choice of the Feynman propagator. All these subtleties put traditional particle-based techniques in danger, and it was recognized early on\footnote{See Section 6 of Dyson's famous ``Missed opportunities'' essay \cite{Dyson:1972sd}.} that the AQFT framework extended to general curved backgrounds must be the right way to proceed.\footnote{Another plausible approach,---Euclidean path integrals,---will be discussed later.} By the 80s, some version of such an approach was available \cite{Ashtekar:1975zn,Dimock1980Alg,Dimock1982Dir,Sewell:1982zz,Kay:1978yp,Kay:1985zs}, but it had shortcomings: it could only describe free fields; there were also problems in subtracting singularities\footnote{Singularity structure of the two-point functions was studied in \cite{Fulling:1978ht,Fulling:1981cf}.} when renormalizing composite operators, such as the stress-energy tensor \cite{Wald94} or general Wick polynomials, where the answer depended on the choice of a reference quasi-free Hadamard state. This prevented both analyzing backreaction and building consistent perturbation theory in the interacting case. Imposing locality and covariance \cite{Wald94} would eventually help to fix these issues. Real progress, however, began in the 90s when it became clear that the microlocal analysis gave a more refined control over the singularities of distributions and allowed to overcome these issues in a more systematic way \cite{Radz92,Radzikowski:1996pa,Radzikowski:1996ei}. The works \cite{Brunetti:1995rf,Brunetti:1999jn,Brunetti:2001dx,Hollands:2001nf,Hollands:2001fb,Sahlmann:2000zr,Strohmaier:2002mm,Hollands:2002ux,Hollands:2004yh,Sanders2010EquivalenceOT} studied the microlocal aspects, in particular: Formulated the microlocal spectral condition, developed a proper (local and covariant) notion of Wick polynomials, including constructions of the covariantly conserved stress tensor, and reduced the renormalization ambiguity to that generated by local gravity counterterms. The gravity counterterms (and more general background counterterms) are generic in the discussion of QFT on classical backgrounds, they lead to fundamental ambiguities and regularization scheme-dependencies that will be mentioned later. QFT in curved space has been a subject of books, monographs and reviews \cite{DeWitt:1975ys,Birrell:1982ix,Fulling:1989nb,BuchOdiSha92,Wald94,Ford97,ParkerToms09,BarFred09,Hollands:2014eia}, in particular see a recent accessible introduction \cite{Witten:2021jzq} (and the follow-up \cite{Witten:2021unn}). In this context one usually talks about globally hyperbolic spacetimes\footnote{A pseudo-Riemannian spacetime (of Lorentzian signature) is globally hyperbolic if it has no closed causal curves, and for any two points, the intersection of the causal past of one with the causal future of the other is compact.} (see \cite{Kay:1992es} for a departure from this condition). The AQFT axioms on globally hyperbolic spacetimes were formulated in \cite{Brunetti:2001dx,Brunetti:2009pn,FewsVerch2015}, see also reviews \cite{Fredenhagen:2004yi,Wald:2006ty,Wald:2009uh,BarFred09,Fredenhagen:2014lda,Fewster:2015nra}, in particular \cite{Wald:2006ty,Wald:2009uh} for the history briefly outlined above, and the review \cite{Fredenhagen:2014lda} and the collection \cite{BarFred09} for more technical details. These axioms are often referred to as \emph{locally covariant quantum field theory} (LCQFT). They are similar to the Haag-Kastler axioms, yet have important differences. Again, there is a net of $C^*$-algebras, but not just on a single globally hyperbolic $M$ and its causal globally hyperbolic subsets; instead, it is defined on all globally-hyperbolic $d$-manifolds simultaneously, with a natural local covariance property with respect to isometric embeddings.\footnote{Further readings on the principle of ``same physics in all spacetimes'' (SPASs) include \cite{Fewster:2011pe,Fewster:2011pn,Fewster:2011ib,Fewster:2015nra}.} Clearly, the Poincare covariance is dropped, and other conditions present in the Haag-Kastler system are replaced by their locally covariant analogs. In \cite{Guido:1999xu,Verch:2001bv,Brunetti:2005gw} superselection sectors and the spin-statistics on curved spaces were considered, for further aspects of the theory see: \cite{Fewster:2012yc,Fewster:2013lqa,Fewster2014PureQS,Brum:2013bia,Fewster:2014gxa,Becker:2014tla,Fewster2015TheSP,Ferguson:2012nd,Sanders2008OnTR}. An alternative approach to AQFT formalizing the OPE on curved spacetime is presented in \cite{Hollands:2009bke}. Constructions of concrete interacting models proceed via the perturbation theory and renormalization, see the original papers and the reviews \cite{Brunetti:1999jn,Hollands:2001nf,Hollands:2002ux,Hollands:2004yh,Brunetti:2009qc,Khavkine:2014zsa,Fredenhagen:2012sb,Fredenhagen:2015iia,Rejz-2016}. Note also a construction of quantum Yang-Mills (YM) as the perturbative AQFT in \cite{Hollands:2007zg}, see also \cite{Fredenhagen2011BatalinVilkoviskyFI}. Other references on gauge theories include \cite{Dimock:1992ff,Pfenning2009QuantizationOT,Dappiaggi:2011zs,Dappiaggi:2011cj,Sanders:2012sf,Hack:2012dm}.

\paragraph{Dynamical C$^*$ algebras.} A novel C$^*$-algebraic approach to QFT is being developed in the recent series of papers \cite{Buchholz:2019rem,BUCHHOLZ2020150,Buchholz:2019mli,Buchholz:2020pmh,Brunetti:2021wev,Brunetti:2021wia,Brunetti:2022cml,Brunetti:2022itx}. It is based on the Lagrangian formulation of field theory, and could probably be called \emph{constructive AQFT}. Indeed, given a Lagrangian $L$, this approach produces a concrete C$^*$-algebra $\cA_L$ called the dynamical C$^*$-algebra in this context. The output obeys the Haag-Kastler axioms and at the same time incorporates ideas from perturbative QFT.

\paragraph{Homotopical AQFT.} Perturbative gauge theories live in the topologically trivial sector. Inclusion of the topological effects like instantons, however, breaks some the axioms of LCQFT: The isotony is violated, as well as the ability to reconstruct global algebras from the local ones.\footnote{We did not describe the latter property in this review, but it is discussed in most of the references.} This fact is explained, for example, in the talks \cite{SchenTalk1,SchenTalk2,SchenTalk3}, see also \cite{Dappiaggi:2011zs,Fewster2016DynamicalLO}. One way to address this problem replaces the space of gauge orbits (configuration space) by a stack given by the corresponding gauge groupoid (a category, whose objects are the bundle-connection pairs and whose morphisms are gauge equivalences). Correspondingly, the ``quantized algebra of functions on fields'' typical to the usual approach is now replaced by some appropriate homotopy\footnote{In this context, the word ``homotopy'' means that various relations like commutativity or associativity hold up to higher homotopies.} dg-algebra. In such a generalized approach (called by practitioners the \emph{homotopical LCQFT}) one obtains, instead of locally covariant nets of $C^*$-algebras, their homotopic dg-versions. Such structures are currently under investigation, see reviews \cite{Benini:2019uge,Bruinsma:2019icp}, the monograph \cite{DYau2019} and the papers \cite{Benini:2013tra,Benini:2015hta,Benini:2016jfs,Benini:2017zjv,Benini:2017fnn,Benini:2018oeh,Benini:2019hoc,Benini:2020gbr,Benini:2020skc,Carmona2021AlgebraicQF,Bruinsma:2021eis,Anastopoulos:2022cez,Grant-Stuart:2022xdh} in which the subject is being developed (see also \cite{hAQFT}).

\paragraph{Haag duality and DHR.} The global symmetries and their role in AQFT (in particular, superselection sectors) were studied by Doplicher, Haag, and Roberts (DHR) \cite{Doplicher:1969tk,Doplicher:1969kp,Doplicher:1971wk,Doplicher:1973at}. To include gauge theories, a modification of the local QFT rules was proposed \cite{Buchholz:1981fj,Frohlich:1993km,Horuzhy}, suggesting to consider, in addition to the bounded regions in Minkowski space, infinite cones. Another approach is being developed in \cite{Casini:2020rgj,Casini:2021zgr,Benedetti:2021lxj}, where the violation of Haag duality $\mathfrak{A}(U) = \mathfrak{A}(U')'$ \cite{HaagBook,Araki63} is at the heart of issue (these authors also consider generalized symmetries and associated extended operators that are responsible for the breakdown of Haag duality, see also \cite{Brunetti:1992zf,Casini:2019kex}).

\paragraph{Factorization algebras and Euclidean perturbative AQFT.} Another approach to Euclidean perturbative QFT, which spiritually fits into the AQFT philosophy, is that of factorization algebras (FA). The notion of FA goes back to \cite{book:BeiDri,lurie2009classification}. The perturbative renormalization in QFT and formulation via factorization algebras was developed in \cite{book:Cos,book:CG1,book:CG2} (see also \cite{GwThesis}). The idea of FA looks superficially similar to the nets of algebras in AQFT, and indeed \cite{GwiRej2017RelatingNA} made comparison between the FA approach and the perturbative AQFT, concluding that the two are closely related. At least for free theories, they show them to be equivalent. In a later paper \cite{Gwilliam:2022vja} the same authors relate observables in the perturbative AQFT and the FA. A general result of \cite{Benini:2019ujs} (where FA are considered on Lorentzian oriented time-oriented globally hyperbolic spaces) abstractly shows their equivalence, modulo natural hypotheses. Therefore, the likely status of the FA approach is that it provides an alternative viewpoint, and technically quite a different approach to constructing concrete models. Some papers that use this framework include \cite{Willi2017,Gorbounov2016ChiralDO,Gwilliam:2017axm,Elliott:2017ahb,Williams2018RenormalizationFH,Gwilliam:2018lpo,Saberi:2019ghy,Saberi:2019fkq,Gwilliam2019AOE,Gwilliam:2020esr,Elliott:2020jai,Saberi:2020pmw,Rabinovich:2020ugt,Bruegmann:2020dfy,Rab21,Elliott:2021peu}, also \cite{Ginot2010HigherHH,Ginot2013NotesOF,Francis2011ChiralKD,Cliff2019}. It is also suggested by \cite{FactonLab} that this approach has close relations to \cite{Hollands:2008ee}.

\subsection{Atiah-Segal-like approach, or Functorial Field Theory}
Following the Atiyah-Segal's axiomatization of TQFT \cite{Atiyah1988} (and its many successes, e.g., classification of fully extended TQFTs \cite{lurie2009classification}), as well as earlier ideas from the work on CFT \cite{Segal:1987,segal_2004}, G.Segal, in a series of lectures \cite{segal_roles}, proposed another set of axioms that are supposed to define a general Euclidean QFT. Similar axioms have been used by Stolz and Teichner \cite{StoTeich04,Stolz2011SupersymmetricFT}, and, apparently, were also considered by Kontsevich (unpublished). These are sometimes referred to as Functorial Field Theory (FFT) \cite{FFTnLab}, though the name is slightly abused, since locally covariant AQFTs discussed earlier are also defined as functors between the appropriate categories (of globally hyperbolic spaces and $C^*$-algebras). We will nevertheless use the term FFT here for concreteness, but we should mention that some authors \cite{LudeSto21,Grady2020ExtendedFT,Grady2021TheGC} call it geometric field theory because it depends on some geometric data on the spacetime, such as the metric. These latter authors seem to have seriously undertaken the task of developing the geometric FFT ideas, and claim to have a definition and even classification (as in the cobordism hypothesis of \cite{lurie2009classification}) of the fully extended geometric FFTs \cite{Grady2020ExtendedFT,Grady2021TheGC}. Recently, the FFT framework was used by Segal and Kontsevich \cite{Kontsevich:2021dmb}, where the definition of QFT on Riemannian manifolds was extended to ``allowable'' complex metrics (a notion serving as a bridge between the Euclidean and the Lorentzian cases). In general, the FFT philosophy for non-topological QFTs has been gaining momentum in the past decade,\footnote{For example, it is apparent from reading \cite[Section 2]{Freed:2012bs} that the authors of this article think about QFT in terms of the FFT paradigm. One can find more examples like this in the literature.} even though the number of papers devoted to non-topological FFTs is still relatively small.\footnote{On the other hand, the field of topological FFT (often called TQFT or Atiyah-Segal TQFT) is thriving, so we omit talking about it, for the same reason that we had to skip CFTs above: the subfield is simply too huge and deserves a separate review. Unfortunately, the related topic of cohomological field theories also has to be skipped.} We should note that an approach to FFTs on CW-complexes (serving as a discretization of spacetime) was developed in \cite{Cattaneo:2012qu,Cattaneo:2015vsa,Cattaneo:2016hqk,Cattaneo:2017tef}. The main idea of FFT is that the field theory is a functor from the category of geometric bordisms (i.e., decorated by some geometric structure) to the category of topological vector spaces (see \cite{Kontsevich:2021dmb}). The functoriality here encodes the gluing axiom that follows from the locality and means that spacetime can be glued from pieces, and these pieces talk to each other only through the boundaries. Namely, FFT on each piece produces a state (co)vector in the tensor product of vector spaces assigned to its boundary components, and gluing (at least in the absence of corners) is dove via composing vectors and covectors. In essence, this is the very same Markov property of Euclidean path integrals that was noticed by Nelson in the 70s \cite{NELSON197397}, as we discussed earlier (see also \cite{DIM2004}). The relation between FFT and AQFT and how the former implies the latter is proposed in \cite{Schr2009}. See also a discussion of physics and formal properties of the gluing axiom in \cite{Dedushenko:2018aox}.

\subsection{Conformal Field Theory (CFT)} 

CFTs form a special subclass of QFTs as they correspond to fixed points of the RG flows. Due to the constraints of conformal symmetry, the operator product expansion (OPE) of their local observables becomes much more concrete and tangible than in generic QFTs. Axiomatically, we could start with any of the approaches reviewed above and specialize them to CFTs. Wightman's and Osterwalder-Schrader axioms in the presence of conformal invariance (really, scale-invariance is enough) are supplied with the OPE relations \emph{under the correlators}. Historically, this has been the most popular approach to CFT, see recent works \cite{Kravchuk:2020scc,Kravchuk:2021kwe} reviewing, among other things, Euclidean CFT axioms and their relation to (W) and (OS) axioms. Functorial QFT in the presence of conformal symmetry in two dimensions leads to the definition of conformal field theory by Segal \cite{Segal:1987,segal_2004}, which actually predates the FQFT axioms. Finally, AQFT axioms supplied by conformal invariance in 2D lead to the notion of conformal nets \cite{Gabbiani:1992ar,Longo:2001zu,Kawahigashi:2002px,Kawahigashi:2003gi,Kawahigashi:2003js,Carpi:2012va,Carpi:2012yf,Carpi:2015fga} (they are known to be related to Vertex Operator Algebras (VOAs), see \cite{Carpi:2015fga,Carpi_2017,TENER2019488,Tener:2018zbp}, the last two of which also mention the relation to FQFT), see also series of works \cite{Bartels:2009ts,BarDouHen1,BarDouHen2,Bartels2013ConformalNI,Bartels2016ConformalNI,Bartels2022ConformalNV}, \cite{HenLect} and \cite{CongNetnLab}.
	
Most results on CFTs appear in the physics literature, but they are often mathematically rigorous (or there are no conceptual obstacles to making them rigorous). The CFTs are usually characterized by the spectrum of local observables and their OPE data, and in two-dimensional spacetime, the enhanced Virasoro symmetry often affords exact solutions \cite{Belavin:1984vu}, connecting to the theory of Vertex Operator Algebras (VOAs). Higher-dimensional CFTs are a subject of an active subfield reviewed in a separate Snowmass paper \cite{Hartman:2022zik}, and there is another one reviewing some aspects of the VOAs \cite{Harrison:2022zee}. Here, we only briefly scratched the surface of the subject, mostly because the CFT literature is really vast and cannot be given any justice in this review.% Still, we should mention a few other highlights: G.Segal's definition of 2d CFT \cite{Segal:1987,segal_2004} (see also its refinement \cite{BLUTE2007101}), Moore-Seiberg analysis \cite{Moore:1988qv}, FFRS construction on TQFT/RCFT \cite{Fuchs:2002cm,Fuchs:2003id,Fuchs:2004dz,Fuchs:2004xi,Fjelstad:2005ua,Fjelstad:2006aw,Runkel:2005qw,Kong2011ConformalFT}. Also, AQFT practitioners study conformal nets of algebras, see. e.g., \cite{Gabbiani:1992ar,Kawahigashi:2002px,Carpi:2012va,Carpi:2015fga} or various references collected at \cite{CongNetnLab}. Unfortunately, we cannot dive more into the enormous conformal literature.

\section{Discussion}
As is apparent from this review and an inevitably incomplete yet huge list of references, the amount of intellectual resources invested into understanding QFT is enormous. Despite that, it is also clear that we are still lacking a single satisfactory unifying viewpoint on the subject. To some extent, the LCQFT axioms of Brunetti-Fredenhagen-Verch-Fewster and the FFT axioms of the last section present the most general and advanced attempts to axiomatize QFT, but even the oldest Wightman axioms still play role in the modern literature (see, for example, \cite{Kravchuk:2020scc,Kravchuk:2021kwe}). There are, however, some obvious issues with these axioms:
\begin{itemize}
	\item The fact that LCQFT faces difficulties in gauge theories and has to be replaced by homotopic AQFT teaches us something. Over the past decades we have learned about dualities in field theories, and understood that ``being a gauge theory'' is not an intrinsic property of a QFT but merely a construction. Indeed, there are known cases when a gauge theory admits a dual non-gauge formulation. Therefore, a model-independent formalism such as AQFT should not treat gauge theories separately. In fact, topological effects occur not only in gauge theories. This suggests that perhaps homotopic AQFT is the right arena for general AQFT machinery, not only gauge theories (it goes in line with the derived mathematics playing more and more role in physics, starting, perhaps, with the Batalin-Vilkovisky formalism). On the other hand, some progress on the issue of gauge theories is being made in \cite{Casini:2020rgj,Casini:2021zgr}.
	\item One can ask a few obvious questions about the FFT approach. It implants the notion of spaces of states, essentially, into the axioms, while the AQFT paradigm emphasizes that the Hilbert space of states is a secondary object not part of the axioms. Furthermore, in case the spatial slices are non-compact, as was emphasized recently in \cite{Witten:2021jzq}, the Hilbert space does not even have to exist. Of course one may overcome this in FFT by only allowing compact spatial slices, but the situation seems a bit uncomfortable.
	\item Another issue that does not seem to be addressed in the FFT framework are ambiguities. As we mentioned in the text, QFTs on generic backgrounds have ambiguities due to the background counterterms that render partition functions regularization-dependent. In some cases, in the presence of extra symmetries, such ambiguities make partition functions valued in bundles, like the $S^2$ partition function in 2d $(2,2)$ SCFTs, which is valued in the K{\"a}hler bundle \cite{Gerchkovitz:2014gta} over the moduli space. In more generic cases (like 2d theories with $(1,1)$ SUSY or less), such an interpretation is lost and the partition function appears completely ambiguous. Thus it might be too naive to assume that the FFT functor always produces a unique answer, in particular always assigns a complex number to a closed spacetime. However, this might be just a normalization issue.
	\item Currently available axiomatic approaches to non-topological QFT do not take extended operators and defects very seriously. That is not say it is impossible: One can include extended operators in the nets of local observables, and it is possible to include defects by modifying the algebras assigned to regions that intersect the defect.\footnote{We thank O.~Gwilliam for a discussion on this point.} One can also incorporate all sorts of (extended or not) observables in the FFT formalism by excising a tubular neighborhood of the observable and assigning the corresponding state to the boundary. However, these issues do not appear to be particularly explored. Even less understood and more mysterious is the case of corners (i.e., extending the theory to codimension $\geq 2$ in the non-topological case).
\end{itemize}
Additionally, we should note that there exists a philosophy typical to the condensed matter literature that QFT describes small perturbations around a critical point of some lattice, many-body or other finite system. We did not include this in the main text as it does not provide a system of axioms. It is not very clear how to relate such a philosophy to any of the axiomatic approaches we have, especially to the AQFT. For example, a lattice system usually comes with a well-defined unique Hilbert space, while the QFT that should emerge from it must, somehow, lose this property. These are of course old questions, some of them have been partially answered in the Constructive Field Theory program for concrete models. We also note a recent increased interest in the lattice approach to QFTs, in particular papers \cite{Radicevic:2021ykf,Radicevic:2021zcz,Radicevic:2021wty}, where a specially designed continuum limit is supposed to address the  above questions.

Besides the issues mentioned before, a number of QFTs studied in the modern literature do not fit into any of the axiom systems currently available.
QFT was originally introduced to marry quantum mechanics with special relativity, but today we know that this was more of a historic accident.
For instance, QFTs exist outside the Lorentz-invariant setting, examples including Lifshitz field theories (see a review \cite{Alexandre:2011kr} and references therein), Horava gravity \cite{Horava:2009uw} (which, however, is a \emph{gravitational} theory,) and many others.
Surely, when placed on curved spaces, such theories are expected to obey some modified version of local covariance, if any at all. 
Hence they do not fit into any of the axiom systems described above.%\footnote{See also \cite{Morgan:2021jpa} for a proposal to break linearity in Wightman's axioms, which, however, we were not able to grasp, and include here for completeness, so the readers could judge themselves on the validity of such a proposal.}

More generally, it has been recently appreciated in the hep-th community \cite{STalk} that our understanding of QFT is incomplete. The standard techniques are very limited, and a number of physically acceptable theories do not fit the old profile of QFT. Such theories include field theories on non-commutative spaces, little string theories, and various exotic theories such as those from \cite{Seiberg:2019vrp,Seiberg:2020wsg,Gorantla:2020xap,Gorantla:2021bda} and references therein. Combining everything we said, there is a clear problem: we do not have the general definition of QFT.

While it is not known how to generalize the notion of QFT yet, one idea is worth mentioning. In \cite{LosevHu} A.Losev and S.Hu made a bold proposal that one should modify the geometry on which the QFT is defined. Instead of working on ordinary manifolds, Riemannian or Lorentzian, one should consider a certain generalization that captures the algebraic operations that are used in constructing QFTs. The authors of \cite{LosevHu} coined the name ``Feynmann geometry,'' and suggested that it should be described by an $A_\infty$-algebra with trace-class operations (such a definition covers many UV regulators: momentum cut-offs, lattices, non-commutativity). In this respect, one should also mention the work of Kontsevich-Soibelman on the $A_\infty$ approach to non-commutative geometry \cite{Kontsevich:2006jb}, which perhaps can be of use. It is also possible that the correct notion of ``Feynman geometry'' should be even more general to cover all the instances of exotic QFTs, if this is the right approach.

\setlength{\unitlength}{1mm}

\newpage

\bibliographystyle{utphys}
\bibliography{QFTax}
\end{document}